\documentclass[aps,pra,reprint,superscriptaddress,longbibliography]{revtex4-2}
\usepackage{orcidlink} 
\usepackage{graphicx}% Include figure files
\usepackage{dcolumn}% Align table columns on decimal point
\usepackage{bm}% bold math
\usepackage{amsthm}
\usepackage{multibib}

\usepackage{braket}
\usepackage{amsmath}
\usepackage{bibentry}
\usepackage{color,soul}

\newcommand{\nocontentsline}[3]{}
\newcommand{\tocless}[3]{\bgroup\let\addcontentsline=\nocontentsline#1{#3}\egroup}

\AtBeginDocument{\let\oldcontentsline\contentsline}
\newcommand{\notoccontentsline}[4]{\oldcontentsline{}{}{}{}}
\newcommand{\droptocpage}{\addtocontents{toc}{\let\protect\contentsline\protect\notoccontentsline}}
\newcommand{\incltocpage}{\addtocontents{toc}{\let\protect\contentsline\protect\oldcontentsline}}

\begin{document}

\preprint{APS/123-QED}
\title{Spatial Optical Simulator for Classical Statistical Models} 
\author{Song-Tao Yu}
\affiliation{Hefei National Research Center for Physical Sciences at the Microscale and School of Physical Sciences, University of Science and Technology of China, Hefei 230026, China}
 
\author{Ming-Gen He}
\affiliation{Hefei National Research Center for Physical Sciences at the Microscale and School of Physical Sciences, University of Science and Technology of China, Hefei 230026, China}
 
\author{Sheng Fang}
\affiliation{Hefei National Research Center for Physical Sciences at the Microscale and School of Physical Sciences, University of Science and Technology of China, Hefei 230026, China}
\affiliation{School of Systems Science and Institute of Nonequilibrium Systems, Beijing Normal University, Beijing 100875, China}

\author{Youjin Deng}
\affiliation{Hefei National Research Center for Physical Sciences at the Microscale and School of Physical Sciences, University of Science and Technology of China, Hefei 230026, China}
\affiliation{CAS Center for Excellence in Quantum Information and Quantum Physics, University of Science and Technology of China, Hefei 230026, China}
\affiliation{Hefei National Laboratory, University of Science and Technology of China, Hefei 230088, China}
 
\author{Zhen-Sheng Yuan}
\affiliation{Hefei National Research Center for Physical Sciences at the Microscale and School of Physical Sciences, University of Science and Technology of China, Hefei 230026, China}
\affiliation{CAS Center for Excellence in Quantum Information and Quantum Physics, University of Science and Technology of China, Hefei 230026, China}
\affiliation{Hefei National Laboratory, University of Science and Technology of China, Hefei 230088, China}


\begin{abstract}
Optical simulators for the Ising model have demonstrated great promise for solving challenging problems in physics and beyond. Here, we develop a spatial optical simulator for a variety of classical statistical systems, including the clock, $XY$, Potts, and Heisenberg models, utilizing a digital micromirror device composed of a large number of tiny mirrors. Spins, with desired amplitudes or phases of the statistical models, are precisely encoded by a patch of mirrors with a superpixel approach. Then, by modulating the light field in a sequence of designed patterns, the spin-spin interaction is realized in such a way that the Hamiltonian symmetries are preserved. We successfully simulate statistical systems on a fully connected network, with ferromagnetic or Mattis-type random interactions, and observe the corresponding phase transitions between the paramagnetic, and the ferromagnetic or spin-glass phases. Our results largely extend the research scope of spatial optical simulators and their versatile applications.
%\begin{description}
%You may use the \texttt{description} environment to structure your abstract;use the optional argument of the \verb+\item+ command to give the category of each item. 
%\end{description}
\end{abstract}

\maketitle
\droptocpage
%\tableofcontents

%%%%%%%%%%%%%%%%%%%%%%%%%%  body  %%%%%%%%%%%%%%%%%%%%%%%%%%
%\section{Introduction}
Optical computing, utilizing the propagation of light for computation,
is an innovative computing architecture,
which exhibits several distinguished features such as high parallelism, high bandwidth, and low power consumption 
when compared to digital computers \cite{caulfield2010future, zhou2022photonic, mcmahon2023physics}. 
In the past decades, optical computing has been demonstrated to have 
a great promise for a variety of widespread applications, 
such as complex physics \cite{ghofraniha2015experimental}, 
neural networks \cite{vandoorne2014experimental,xu202111,lin2018all,feldmann2019all,wetzstein2020inference}, 
photonic programmable signal processors \cite{zhuang2015programmable,perez2017multipurpose,bogaerts2020programmable},  
cryptography \cite{pai2023experimental} and so on.

A remarkable achievement in the research of optical computing is the experimental realization 
of optical simulators for the Ising model~\cite{Marandi,Inagaki,Okawachi,Roques,Pierangeli},
which is perhaps the most fundamental model in statistical physics. 
These simulators are often termed photonic Ising machines, 
with their scalability greatly enhanced by encoding spins using liquid crystal spatial 
light modulator (LC-SLM) \cite{Pierangeli,yamashita2023low,pierangeli2020adiabatic}. 
Very recently, the digital micromirror device (DMD) has also been used, 
where each Ising spin is encoded by a $2 \times 2$ unit of tiny mirrors 
in the DMD~\cite{leonetti2021optical}. The photonic Ising machine based on LC-SLM 
or DMD is commonly referred to as the spatial photonic Ising machine (SPIM), 
which can reduce the computing complexity from $O(N^2)$ to $O(N)$ for the annealing algorithm 
in fully connected models with $N$ spins \cite{pierangeli2020adiabatic,pierangeli2021scalable}. 
It has also been demonstrated that SPIM can provide a powerful platform for studying phase transitions \cite{pierangeli2021scalable,kumar2023observation,leonetti2021optical,fang2021experimental} and solving combinatorial optimization problems \cite{mohseni2022ising,prabhakar2023optimization}.

Besides the Ising model, there exist many other important classical systems in statistical and condensed-matter physics, among which typical examples include the clock, $XY$, Potts, and Heisenberg models. The spins of these systems assume discrete or continuous values, and the Hamiltonians possess various symmetries like $Z_q$, $U(1)$, $S_q$, and $O(3)$. These models play a central role in the modern theory of phase transitions and have broad applications in optimization problems \cite{zhang2006complex,so2007approximating,frieze1997improved,lewis2015guide}. Extensive effort has been devoted to simulating these statistical models by optical or optoelectronic systems that include optical parametric oscillators \cite{hamerly2016topological,takeda2017boltzmann, Honari, inaba2022potts},   exciton-polaritons \cite{RN33,RN6,Kalinin,harrison2022solving} and coupled lasers \cite{nixon2013observing,gershenzon2020exact}. However, these optical systems face challenges in handling large-scale spin Hamiltonians. Fortunately, owing to the excellent scalability, the SPIM-type simulators are capable of effectively addressing these challenges \cite{Pierangeli,pierangeli2021scalable}. 
 
\begin{figure}
\includegraphics[width=0.98\columnwidth]{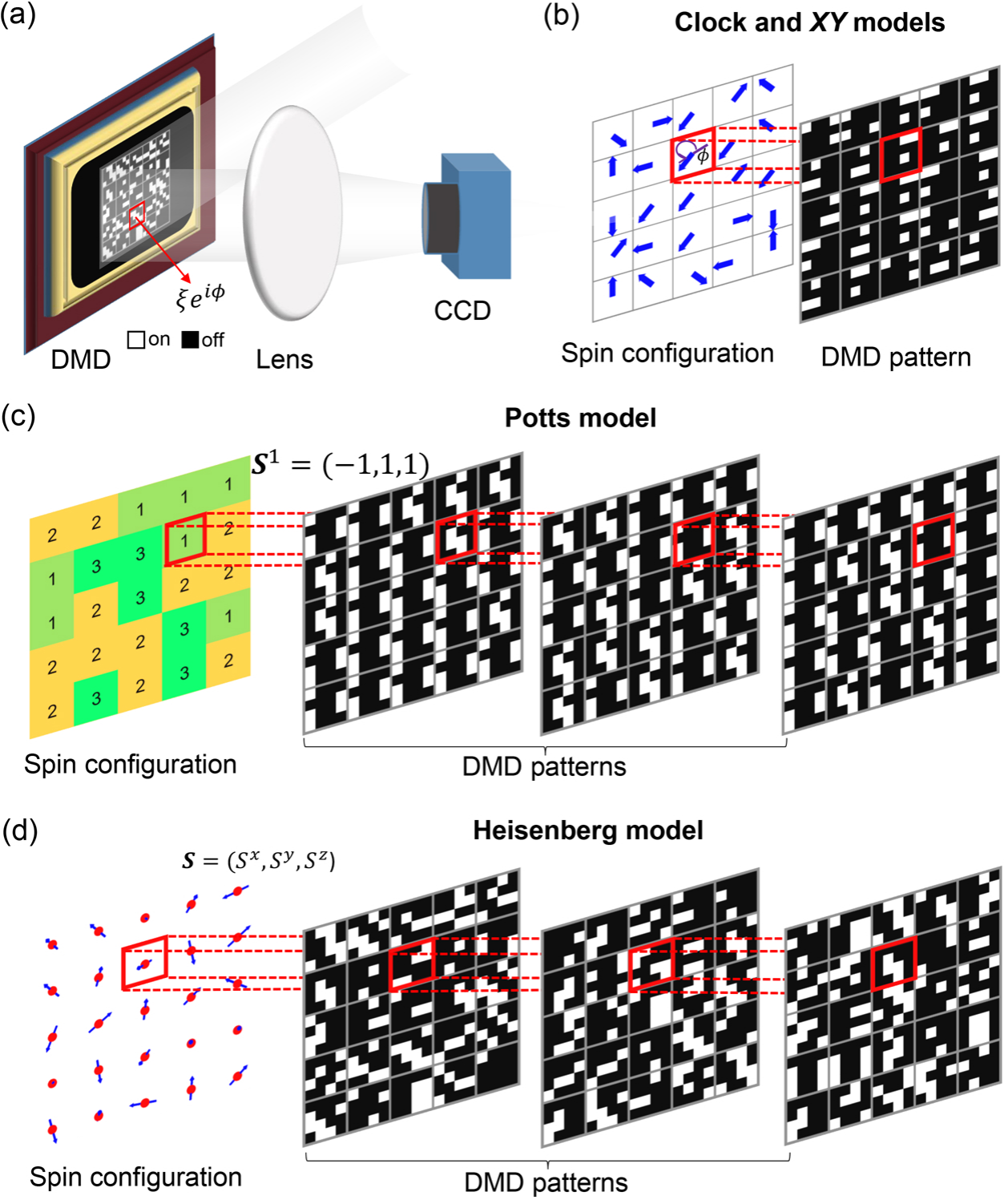}% Here is how to import EPS art
\caption{\label{setup} Scheme of the DMD-based optical simulators for statistical models.
(a) Experimental setup of the spatial optical simulator. 
The spin configuration is mapped onto the DMD patterns, and 
the corresponding energy can be obtained from the detected intensity image on the CCD camera. 
(b) The spins of the $q$-state clock model can be encoded by modulating the phase of the light field,
and the $q \to \infty$ limit gives the $XY$ model.
(c) Each spin of the $3$-state Potts model is encoded by three superpixels 
such that the Kronecker delta function for the pairwise interaction is expressed 
as vector-vector multiplication. 
(d) Each spin component of the Heisenberg spin, $\bm{S}=(S^x,S^y,S^z)$,
 is encoded by a superpixel with the constraint $|\bm{S}|=1$.
}
\end{figure}

In this Letter, we develop a SPIM-type optical simulator for a series of classical statistical models based on a single DMD device. Compared to the LC-SLM technology that suffers from low refresh rates, significant temporal-phase fluctuations, and crosstalk between neighboring pixels, the DMD offers faster speed and greater stability~\cite{turtaev2017comparison}. Utilizing the fact that a superpixel is constructed from  $4 \times 4$ pixels, each of which is either in an ``on'' or ``off'' state, we use the superpixel to encode the amplitude or phase of spins in statistical models. Further, to preserve the symmetries in the Hamiltonians, we adopt a general strategy to engineer the spin-spin interactions by modulating one or more sequences of light patterns. The DMD, the CCD camera, and the digital computer are then integrated into a feedback system to create the simulator.

As applications, we apply the DMD-based optical simulator to study the clock, $XY$, Potts, and Heisenberg models on a fully connected network with ferromagnetic or Mattis-type random interactions, and successfully observe the phase transitions of these systems. The spontaneous symmetry breaking is clearly demonstrated in the evolution of the probability distribution for the order parameter, revealing the number of ferromagnetic ground states or the replica-symmetry-breaking nature of the spin-glass phase. Our experimental results agree quantitatively with Monte Carlo (MC) simulations on digital computers. 

\textit{Statistical models.}%
---%
Given a network, the Hamiltonians of the statistical models read as 
\begin{eqnarray}
	\label{eq:models}
	{\cal H}_\text{clock} &=& -\sum_{mn} J_{mn} \cos (\phi_m-\phi_n) \;,  \hspace{1.5mm} 
	\text{clock or $XY$} \nonumber  \\ 
	{\cal H}_\text{Potts} &=&  -\sum_{mn} J_{mn} \delta_{\sigma_m, \sigma_n} \;,  
	\hspace{13mm} \text{Potts}  \\
	{\cal H}_\text{O(n)} &=& -\sum_{mn} J_{mn} {\bm S}_m \cdot {\bm S}_n  \;, 
	\hspace{11mm}  \text{Heisenberg}  \nonumber 
\end{eqnarray}
where $J_{mn}$ represents the coupling strength between sites $m$ and $n$, and 
the summation is over all pairs of interacting spins. 
For the $q$-state clock and $q$-state Potts, the spins take discrete values as
 $\phi =  2\pi k /q$ with $k \in \{1,\cdots,q \}$ and $\sigma \in \{1,2,\cdots,q \}$,
respectively. The $q=2$ case corresponds to the Ising model,
and the $q \to \infty$ clock model becomes the $XY$ model.
The Ising, $XY$, and Heisenberg models can also be included within 
the O($n$) spin model, in which each spin is an $n$-dimensional vector of unit 
length, i.e., $|{\bm S}|=1$. 
Despite the relatively simple form of the Hamiltonian, 
these models exhibit very rich physics partly due to their wide range of symmetries.

\textit{Optical encoding of statistical models.}%
---%
The architecture of the spatial optical simulator is depicted in Fig.~\ref{setup}(a). The DMD (DLP 3000, Texas Instruments) consists of $608\times684$ tiny mirrors (called pixels), each with a pitch size of $d$ = 7.64 $\mu$m. Each mirror can be switched between a $+12^\circ$ ``on'' and a $-12^\circ$ ``off'' orientation. 
The images loaded onto the DMD are then binary images, with values of 1 (white pixel) or 0 (black pixel), 
respectively for the ``on'' and ``off'' states [Fig.~\ref{setup}(a)]. 
The DMD employs a method called ``superpixel'' to achieve spatial amplitude and phase modulation 
of the light field \cite{goorden2014superpixel}. 
As shown by the red area in Fig.~\ref{setup}(a), a superpixel is constructed 
by $4\times 4$ DMD pixels such that it can be manipulated to 
have a tunable spatial amplitude and phase modulation of the light field, 
denoted as $\xi e^{i\phi}$, with $\xi$ ranging between 0 and 1 and $\phi$ 
varying from 0 to $2\pi$; namely, a superpixel can be programmed to encode 
an arbitrary complex number. 
As the binary number system for the von Neumann computer, 
any classical spins, irrespective of being discrete or continuous, 
can be encoded by using one or more superpixels. 

We illuminate the DMD with light at a wavelength of 785 nm. 
The modulated light field reflected from DMD propagates through a lens (focal length $f$ = 200 mm), 
and the intensity at the center position, detected by a CCD camera (DCU224M-GL, Thorlabs) 
at the back focal plane, can be written as
\begin{equation}
I=\sum_{mn}\xi_m \xi_n e^{i(\phi_m-\phi_n)} \; , \label{Intensity}
\end{equation}
where $m$ and $n$ are indices of the superpixels. 
For a fully connected network, one can further show that, with Eq.~(\ref{Intensity}), 
the energy of any spin configuration for these statistical models 
can be directly read out from the camera.

For the $q$-state clock model, the spin, taking value as $\phi = 2\pi k/q$, 
can be represented by the light-field phase 
of a single superpixel as illustrated in Fig.~\ref{setup}(b). 
Further, by rewriting Eq.~(\ref{Intensity}) as 
$I = \sum_{m>n}2\xi_m \xi_n \cos(\phi_m-\phi_n)+ \rm{const}$ 
and setting ${\cal H}_{\rm clock}=-I$, one can see 
that the cosine-function interaction in Eq.~(\ref{eq:models}) is simply 
realized with the coupling strength $J_{mn} = \xi_m \xi_n$, 
and that the energy of a spin configuration is just the intensity of the center 
position on the detection plane (apart from an unimportant constant). 
Thus, a spin configuration can be optically encoded by precisely 
tuning the light-field phase of each superpixel to its target value,
and the set of coupling strength $J_{mn}$ can be realized 
by controlling the light-field amplitudes.

The $q \to \infty$  clock model reduces to the $XY$ model. In practice, a $4\times 4$ superpixel in the current DMD technology has only 6561 discrete fields \cite{goorden2014superpixel}. Since a single superpixel is used to encode a clock spin, some discretization effects arise in the $XY$ model. The superpixel method for phase modulation can achieve more than a hundred gray-scale levels, allowing the value of $q$ to reach the hundreds. Nevertheless, it is theoretically known that, for $q \geq 5$, the phase transition of the clock model is already in the $XY$ universality class \cite{borisenko2011numerical,kumano2013response,chen2022monte}.

For the $q$-state Potts model, the spin also takes a discrete, but the interaction 
is of the Kronecker delta function and the Hamiltonian has the $S_q$ symmetry \cite{wu1982potts}.
We embed the Potts spin in a $q$-dimensional space 
and represent each  Potts state, $\sigma \in \{1,\cdots,q \}$, 
by a specific vector ${\bm S}^\sigma$ such that the Cartesian coordinate 
of ${\bm S}^\sigma$ is $-1$ along the $\sigma$th axes and, otherwise, $+1$.
For instance, the Potts spins for $q=3$ are 
${\bm S}^1 = (-1,1,1)$, ${\bm S}^2 = (1,-1,1)$ and ${\bm S}^3 = (1,1,-1)$.
It can be checked that the delta-function interaction in Eq.~(\ref{eq:models}) 
can be rewritten as the vector multiplication, i.e., 
$\delta_{\sigma_m, \sigma_n} = ({\bm S}_m \cdot {\bm S}_n -q+4)/4$, 
and, thus, the Hamiltonian symmetry is preserved.

In experiments, the light field reflected from each superpixel on the DMD is precisely 
tuned to have phase $ \phi =0$ or $\pi$, representing the $+1$ or $-1$ states, respectively. 
For each spin configuration, we sequentially control DMD to display $q$ images [Fig.~\ref{setup}(c)]
and measure their corresponding far-field diffraction intensity at the center, denoted as  $I_k$, using the CCD camera. 
According to Eq.~(\ref{Intensity}), the total light intensity is 
$I = \sum_k I_k = \sum_{mn} \xi_m \xi_n {\bm S}_m \cdot {\bm S}_n$, 
with $\xi$ the light-field amplitude.
As a result, a spatial optical simulator is successfully constructed for 
the $q$-state Potts model~[Eq.~(\ref{eq:models})] on the fully connected network, 
where the coupling strength is $J_{mn}=\xi_m \xi_n /4$.

For the Heisenberg model, since the spin is a three-dimensional unit vector, 
${\bm S} = (S^x, S^y, S^z)$ with $|{\bm S}|=1$,
a spin configuration can be also represented by three DMD images [Fig.~\ref{setup}(d)], 
and each superpixel on a DMD image encodes a real coordinate $S^k$ ($k=x,y,z$).
The light-field phase on the superpixel is precisely controlled to 
be $\phi = 0$ or $\pi$, encoding the sign of $S^k$, and the amplitude $\xi$ is 
tuned to match the magnitude $|S^k|$. 
The total light intensity, which corresponds to the Fourier transform of 
the three DMD images consecutively detected by the CCD camera, 
is  $I = \sum_{mn} (S^{x}_{m}S^{x}_{n}+S^{y}_{m}S^{y}_{n}+S^{z}_{m}S^{z}_{n}) =
\sum_{mn} {\bm S}_m \cdot {\bm S}_n$, representing the Heisenberg model 
with homogeneous coupling $J_{mn}=1$. 
An additional DMD image can be employed to encode a set of nonuniform coupling strength 
as $J_{mn}=\xi_m \xi_n$.

These statistical models with coupling strength $J_{mn}=\xi_m \xi_n$ are also known as the Mattis models~\cite{RN43,nishimori2001statistical}. Note that, the complex coupling can be realized through spatial light modulation methods ~\cite{Pierangeli}, optoelectronic correlation computing~\cite{huang2021antiferromagnetic}, and other techniques~\cite{luo2023wavelength,sakabe2023spatial,sun2022quadrature}.  

\begin{figure*}
	\includegraphics[width=0.8\linewidth]{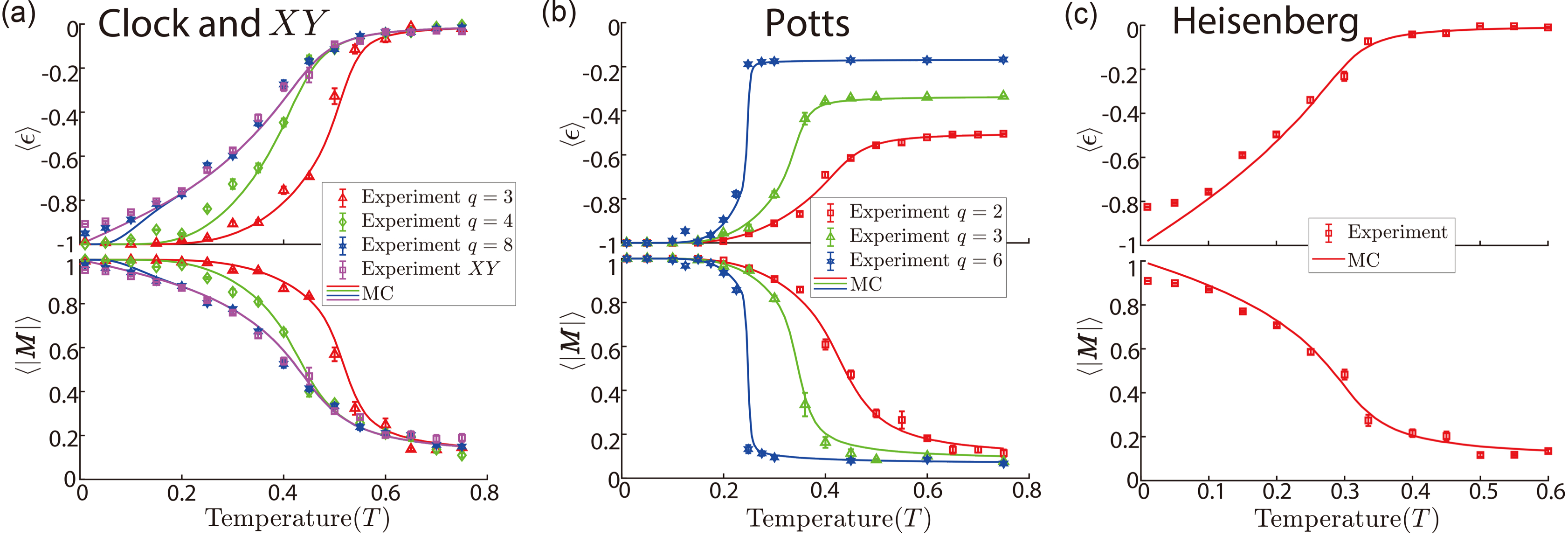}% Here is how to import EPS art
	\caption{\label{energy_magnetization} The energy density $\langle \epsilon \rangle$ and the magnetization $\langle |\bm{M}| \rangle$ of (a) the clock and $XY$, (b) the Potts, and (c) the Heisenberg models as a function of the temperature $T$. The dots and lines show the results of experiments and MC simulations on a digital computer, respectively. 
	}
\end{figure*}

\textit{Optical simulation of statistical models.}%
---%
In practical experiments, the spatial optical simulator functions as a feedback system comprising the DMD, the camera, and the digital computer. Within this setup, we integrate the Metropolis-Hastings algorithm to serve as a Boltzmann distribution sampler for the clock, $XY$, Potts, and Heisenberg models \cite{hastings1970monte}. Specifically, in a given statistical model, the spin configuration is stored in the computer memory [for the Heisenberg model, the normalization is required for each site, i.e., $(S^x)^2+(S^y)^2+(S^z)^2=1$], and, according to the aforementioned coding schemes, it is encoded onto one or more DMD images. Then, the CCD camera is employed to detect the light field reflected by the DMD, and, from the light intensity at the center, the energy $H$ of the spin configuration is obtained. We randomly flip one spin and then measure the energy change before and after the flip. We accept the updated spin configuration with a probability $p=\mathrm{min}\{1,e^{-\Delta H/(2NT)}\}$, where $\Delta H$ represents the variation in energy, $N$  is the number of spins, and $T$ is the effective temperature. Furthermore, to reduce the number of iterations, we employ simulated annealing to generate the ensemble distributions at different temperatures \cite{chibante2010simulated,gelman1995bayesian}. Finally, to avoid the ergodicity problem, we also employ a global update scheme by rotating the $XY$ or Heisenberg spin on each site by a fixed angle (randomly chosen) and permuting the clock or Potts spins. The global update is only performed after every $2N$ spin-flip update for a system of $N$ spins. Although it does not change the energy of the system, it is important to explore the number of ground states at low temperatures. By storing the spin configuration at different temperatures $T$, we can track the evolution of observables as temperature changes. Based on this approach, we study spin models with ferromagnetic and Mattis-type random interactions, considering a system with $N = 100$ spins.
 
In the ferromagnetic models, we investigate the phase transition behavior of the $q=3,4,8$-state clock, $XY$, $q=2,3,6$-state Potts, and Heisenberg models. Theoretical critical temperatures and transition types for these models are presented in Table~\ref{tab:table1}. 
We observe variations in the energy density $\epsilon$ 
and the magnetization $\bm M$ across different temperatures. 
For the clock, $XY$, Potts, and Heisenberg models, the energy density is
\begin{equation}
	\epsilon = -\frac{1}{N(N-1)}\left\{
	\begin{array}{ll}
		\sum_{m\neq n} \cos(\phi_m-\phi_n)  &\text{clock or $XY$} \\
		\sum_{m\neq n} \delta_{\sigma_m,\sigma_n} &\text{Potts}  \\    
		\sum_{m\neq n} \bm{S}_m \cdot \bm{S}_n &\text{Heisenberg}		 
	\end{array},\right.
\end{equation}
and the magnetization is defined as  
\begin{equation}
	\bm{M} =\frac{1}{N} \left\{
	\begin{array}{ll}
		\sum_{n} e^{i\phi_n}  &\text{clock or $XY$}\\
		\sum_{n} \frac{q\delta_{\sigma_n,1}-1}{q-1} &\text{Potts}  \\    
		\sum_{n} \bm{S}_n &\text{Heisenberg}		
	\end{array}.\right.
\end{equation}

Figure~\ref{energy_magnetization} illustrates the changes in the energy density $\langle \epsilon \rangle$ and 
the magnetization $\langle |\bm M| \rangle$ near the phase transition points, 
where the $\langle \dots  \rangle$ denotes the thermal average. 
For the clock and $XY$ model [Fig.~\ref{energy_magnetization}(a)], 
the experimental data (dots) are generally in agreement with the results obtained from MC simulations 
on a digital computer (lines). 
The error bars are calculated taking into account the correlation between samples. 
Notably, for $q=3$, there are significant changes in energy density and magnetization 
near the critical point $T_c$. For $q>3$ and the $XY$ model, the changes appear more gradual. 
These observations suggest different types of phase transitions, 
though they do not conclusively indicate whether the transitions are first-order or continuous. 
Similarly, the Potts model shows distinct behaviors in energy density and magnetization 
near the critical point $T_c$, as presented in Fig.~\ref{energy_magnetization}(b). 
For $q=2$, the changes are more gradual, while for $q>2$, the changes are more abrupt. 
These variations also hint at different types of phase transitions. 
For the Heisenberg model shown in Fig.~\ref{energy_magnetization}(c), 
discrepancies between simulated and experimental values are observed at low temperatures. This discrepancy arises from discretization encoding errors, detection noise, or aberrations. In particular, the discretization encoding errors, due to the limited total number of different fields in a superpixel, are expected to be more pronounced in the Heisenberg model than in the $XY$ model, as observed at sufficiently low temperatures [Fig.~\ref{energy_magnetization}(a)]. Additionally, multiple measurements in the Heisenberg model also introduce more errors.

\begin{table}[h]
	\caption{\label{tab:table1}
		Critical temperatures $T_c$ and transition types for statistical models in theory. The transition types are first-order (1st) and second-order (2nd).}
	\begin{ruledtabular}
		\begin{tabular}{ccccccc}
			\textrm{Model}&
			\textrm{$T_c$}&	
			\parbox{5em}{\centering Transition \\ type}&
			&
			\textrm{Model}&
			\textrm{$T_c$}&
			\parbox{5em}{\centering Transition \\ type}\\
			\colrule
			3-state clock & 0.54 & 1st&&2-state Potts & 0.5 & 2nd\\
			4-state clock & 0.5  & 2nd&&3-state Potts & 0.36& 1st\\
			8-state clock & 0.5  & 2nd&&6-state Potts & 0.25& 1st\\
			$XY$          & 0.5  & 2nd&& Heisenberg & 0.33 & 2nd
		\end{tabular}
	\end{ruledtabular}
\end{table}
\begin{figure}
	\includegraphics[width=0.98\columnwidth]{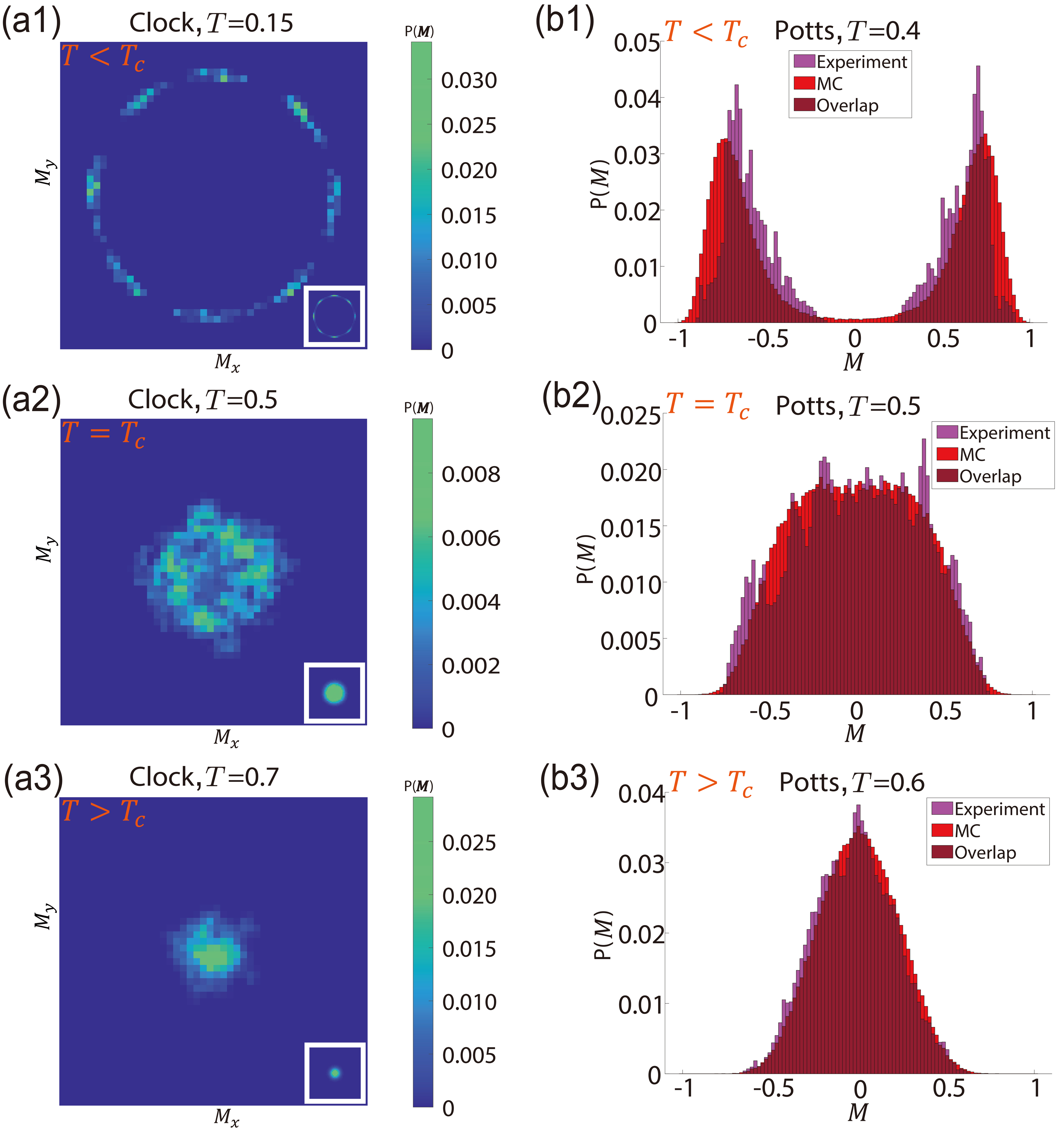}% Here is how to import EPS art
	\caption{\label{m_distribution} The spontaneous symmetry breaking in the 8-state clock and 2-state Potts models. (a1)-(a3) The probability distribution of the magnetization of the 8-state clock model at $T<T_c$, $T=T_c$, and $T>T_c$. The inset shows the MC simulation results on a digital computer. (b1)-(b3) The histogram of the magnetization $\bm{M}$ for the different temperatures in the 2-state Potts model. Magenta and scarlet bars indicate experimental results and the MC simulations on a digital computer respectively. The deviation of the histogram originates from the camera detection noise and encoding errors.}
\end{figure}

We then demonstrate the probability distribution of the order parameter $\bm{M}$ before and after the critical temperature for the clock and Potts models. Figures~\ref{m_distribution}(a1)-(a3) illustrate the phase transitions of the 8-state clock model in experiments and MC simulations on a digital computer (inset). When the temperature is above the critical temperature, the system is in the disordered phase, where the orientation of the spins is randomly distributed, and the magnetization follows a two-dimensional Gaussian distribution. As the temperature decreases, the average magnetization vector aligns in eight directions with equal probability. This process illustrates the phenomenon of spontaneous symmetry breaking. In the case of the 2-state Potts model (Ising model), histograms of the magnetization for $T<T_c$, $T=T_c$ and $T>T_c$ are presented in Figs.~\ref{m_distribution}(b1)-(b3). Both experiments (magenta) and MC simulations on a digital computer (scarlet) show a consistent behavior in the distribution of $\bm{M}$ from a single peak to a double peak as the temperature decreases, which indicates the spontaneous symmetry breaking during the continuous phase transition, i.e., from paramagnetic phase to ferromagnetic phase.

\begin{figure}
	\includegraphics[width=0.98\columnwidth]{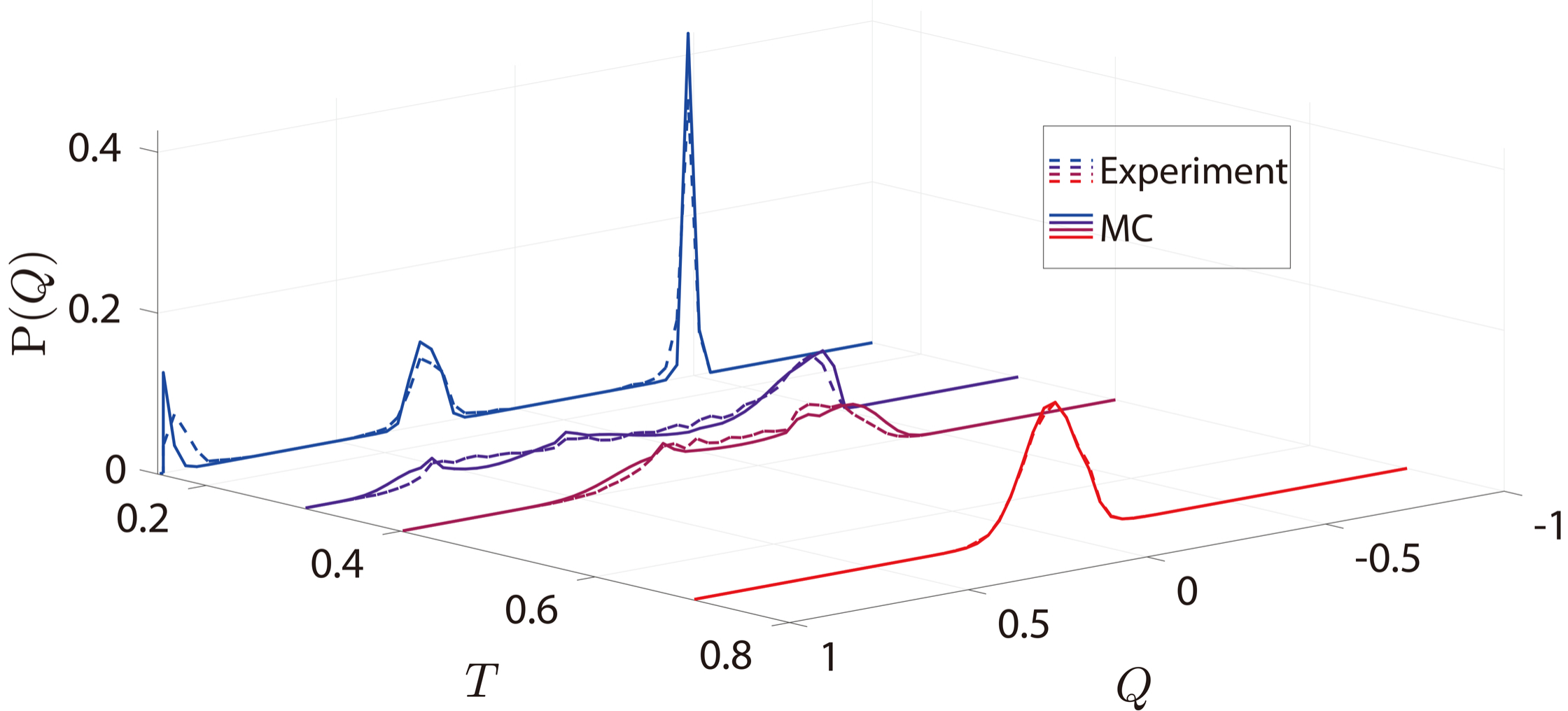}% Here is how to import EPS art
	\caption{\label{spin_glass} The spin glass in the 3-state clock model. The probability distribution of the parameter $Q$ as a function of $T$, with $p = 0.5$. The dashed lines and solid lines represent the results of the experiment and the MC simulations on a digital computer.}
\end{figure}

We also simulate the Mattis-type spin-glass system within the 3-state clock model, where the interactions follow $J_{mn}=\xi_m \xi_n$, with probability distribution ${\rm Pro}(\xi_m)=p\delta_{\xi_m,-1}+(1-p)\delta_{\xi_m,+1}$ and $p=0.5$. We generate 100 independent replicas using a quenched approach, and characterize the similarity between two replicas $\alpha$ and $\beta$ by measuring the replica order parameter $Q_{\alpha\beta}=(1/N)\sum \cos(\phi_m^\alpha-\phi_n^\beta)$. Figure \ref{spin_glass} shows the experimental and MC simulation results for the probability distribution of the parameter $Q_{\alpha\beta}$. At high temperatures, there is no correlation between the replicas, and the distribution of $Q$ shows an unimodal distribution centered around zero. As the temperature decreases, the distribution of $Q$ exhibits multiple peaks, indicating the presence of multiple degenerate ground states and a multivalley energy landscape. These results demonstrate the existence of a spin-glass phase in the system. Finally, we mention that our optical simulator can readily simulate the clock, $XY$, Potts, and Heisenberg models with Mattis-type random interactions, which can exhibit rich phenomena like gauge glass, Potts-spin glass, and recursive phase transitions, etc. 

\textit{Conclusion.}%
---%
In summary, we experimentally realize a DMD-based optical simulator and develop a strategy to simulate a variety of classical statistical models with distinct symmetries. The phase transitions of these systems are observed, and the properties of low-temperature phases are investigated. It is emphasized that the strategy is general, for encoding classical spins and realizing spin-spin interactions, and can be applied to statistical-physics systems beyond the current ones. It may also be used to explore more complex interactions, including many-body interactions~\cite{kumar2023observation}, as well as more general networks~\cite{Pierangeli,huang2021antiferromagnetic,luo2023wavelength,sakabe2023spatial,sun2022quadrature}, providing a promising and powerful tool to achieve scalable simulators to tackle complicated computational tasks. 
%In summary, we experimentally realize a DMD-based optical simulator and develop a strategy to simulate a variety of classical statistical models with distinct symmetries. The phase transitions of these systems are observed, and the properties of low-temperature phases are investigated. It is emphasized that the strategy is general, for encoding classical spins and realizing spin-spin interactions, and can be applied to statistical-physics systems beyond the current ones. It may also be used to explore more complex interactions, including many-body interactions~\cite{kumar2023observation}, as well as more general networks~\cite{Pierangeli,huang2021antiferromagnetic,luo2023wavelength,sakabe2023spatial,sun2022quadrature}. Owing to its high speeds and stability, the DMD technology can provide a promising and powerful tool to achieve scalable simulators to tackle complicated computational tasks. 

\smallskip
\textit{Acknowledgments.}%
---%
	This work was supported by the National Natural Science Foundation of China (No. 12125409, No. 12275263),  
	the Innovation Program for Quantum Science and Technology (No. 2021ZD0302000, No. 2021ZD0301900), 
	the Anhui Initiative in Quantum Information Technologies, 
	and the Natural Science Foundation of Fujian Province of China(No. 2023J02032).

\bibliography{OpticalSimulator_arXiv}%main   Produces the bibliography via BibTeX.

\end{document}